\documentclass[a4paper]{article}

\usepackage{INTERSPEECH2019}
\usepackage{subfigure}
\usepackage{gensymb}
\usepackage{multirow}
\usepackage{xcolor}
\usepackage{hyperref}
\usepackage[section]{placeins}

\DeclareMathOperator{\boldI}{\mathbf{I}}

\graphicspath{ {figures/} }

\title{Regression and Classification for Direction-of-Arrival Estimation with Convolutional Recurrent Neural Networks}
\name{Zhenyu Tang, John D. Kanu, Kevin Hogan, Dinesh Manocha\thanks{This work was supported in part by ARO grant W911NF-18-1-0313 and Intel. Project page \url{https://gamma.umd.edu/pro/speech/doa} }}
\address{
  University of Maryland}
\email{zhy@cs.umd.edu, jdkanu@cs.umd.edu, khogan@cs.umd.edu, dm@cs.umd.edu}

\begin{document}

\maketitle
\begin{abstract}

We present a novel learning-based approach to estimate the direction-of-arrival (DOA) of a sound source using a convolutional recurrent neural network (CRNN) trained via regression on synthetic data and Cartesian labels. We also describe an improved method to generate synthetic data to train the neural network using state-of-the-art sound propagation algorithms that model specular as well as diffuse reflections of sound. We compare our model against three other CRNNs trained using different formulations of the same problem: classification on categorical labels, and regression on spherical coordinate labels. In practice, our model achieves up to 43\% decrease in angular error over prior methods. The use of diffuse reflection results in 34\% and 41\% reduction in angular prediction errors on LOCATA and SOFA datasets, respectively, over prior methods based on image-source methods. Our method results in an additional 3\% error reduction over prior schemes that use classification networks, and we use 36\% fewer network parameters.

\end{abstract}

\section{Introduction}
Estimating the direction-of-arrival (DOA) of sound sources has been an important problem in terms of analyzing multi-channel recordings~\cite{knapp1976generalized, brandstein1997robust}. In these applications, the goal is to predict the azimuth and elevation angles of the sound source relative to the microphone, from a sound clip recorded in any multi-channel setting. One of the simpler problems is the estimation of the DOA on the horizontal plane~\cite{xiao2015learning}. More complex problems include DOA estimation in three-dimensional space or the identification of both direction and distance of an audio source. Even more challenging problems correspond to performing these goals in noisy and reverberant environments.

To analyze spatial information from sound recordings, at least two microphones with known relative positions must be used. In practice, various spatial recording formats including binaural, 5.1-channel, 7.1-channel, etc. have been applied to spatial audio related systems~\cite{zhang2017surround}. The Ambisonics format decomposes a soundfield using a spherical harmonic function basis~\cite{gerzon1973periphony}. Compared with its alternatives, Ambisonics has the advantage of being hardware independent--it does not necessarily encode microphone specifications into the recording.

Recent work~\cite{perotin2018crnn} has applied the Ambisonics format to DOA estimation and trained a CRNN classifier that yields more accurate predictions than a baseline approach using independent component analysis. While a regression formulation seems more natural for the problem of DOA estimation, some recent work~\cite{xiao2015learning} suggests a regression formulation may yield worse performance than that of the classification formulation for multi-layer perceptrons. In this work, we present a novel learning-based approach for estimating DOA of a single sound source from ambisonic audio, building on an existing deep learning framework~\cite{perotin2018crnn}. We present a CRNN which predicts DOA as a 3-D Cartesian vector. We introduce a method to generate synthetic data using geometric sound propagation that models specular and diffuse reflections, which results in up to 43\% error reduction compared with image-source methods. We conduct a four-way comparison between the Cartesian regression network, two classification networks trained with cross-entropy loss, and a regression network trained using angular loss. Finally, we investigate results on two 3rd-party datasets: LOCATA~\cite{LOCATA2018a} and SOFA~\cite{perez2018ambisonics}, where our best model reduces angular prediction error by 43\% compared to prior methods.

Section 2 gives an overview of prior work. We propose our method in Section 3. Section 4 presents our results on two benchmarks and we conclude in Section 5.
\section{Related Work}
\subsection{Overview}
One classic approach to DOA estimation is to first determine the time delay of arrival (TDOA) between microphone array channels, which can be estimated by generalized cross correlation~\cite{knapp1976gcc} or least squares~\cite{huang2001ls}. The DOA can be computed from known TDOA and the array layout directly. Another approach is to use the signal subspace, as in the MUSIC algorithm~\cite{schmidt1986music}. With some restrictions on operating conditions, these techniques are very effective. However, they do not perform well in highly reverberant and noisy environments, or when the placement of signal sources is arbitrary~\cite{dibiase2001robust}. More recently, researchers have applied modern machine learning techniques to speech DOA estimation with the goal of improving performance in noisy, realistic environments, which can be categorized into \emph{classification} and \emph{regression} networks. For both networks, angular error for a single example is proportional to the angular distance between the predicted and actual DOA.

\subsection{Classification Formulations}

In the classification formulation, DOA is encoded using a \emph{categorical} representation: an approximately uniform mesh-grid defines the score for each of a finite set of possible categories, where each category corresponds to a unique region of the continuous DOA space. The mesh grid is defined by subdividing the DOA space at a given resolution. The DOA is decoded as the direction associated with the bin with highest score. The categorical formulation uses a discrete encoding, lending itself to a class-based formulation of DOA estimation. Generalized cross correlation (GCC) feature vectors of a microphone array input have been fed to a multilayer perceptron classifier, which predicts a DOA in one angular dimension~\cite{xiao2015learning} and show superior performance over the classic least square method~\cite{huang2001ls} in both simulated and real rooms of various sizes. Perotin et al.~\cite{perotin2018crnn} calculate acoustic intensity vectors using a first-order Ambisonics representation of audio. This representation serves as input to a CRNN, which predicts a DOA in two angular dimensions. Their CRNN yields more accurate predictions than a baseline approach using independent component analysis. CRNNs have also been used in~\cite{adavanne2018sound} to identify the DOA for overlapping sound sources, in two angular dimensions.

\subsection{Regression Formulations}

In the regression formulation, two representations of DOA are commonly used, which we refer to as \emph{Cartesian} and \emph{spherical}. With the Cartesian representation, DOA is encoded as a three-dimensional vector in Cartesian $(x,y,z)$ coordinates, pointing toward the source. With the spherical representation, DOA is encoded as a two-dimensional vector of azimuth ($\theta$) and elevation ($\phi$) angles. Both formulations encode DOA in continuous space, leading to a regression formulation of DOA estimation. In prior work, regression formulations have not shown superior empirical results for DOA estimation. Higher angular errors for regression than for classification is claimed in~\cite{xiao2015learning}. CNN regression has been used in~\cite{veradiaz2018CNN} to estimate the Cartesian coordinates of a sound source in 3-D space. CRNN regression has been used in~\cite{adavanne2018sound} and higher angular error is observed for DOA estimates for regression than for classification. Similar to this result, our experiments show a higher angular error for regression on spherical DOA than for classification. However, we discover a \emph{lower} angular error for regression on Cartesian DOA, than for both classification on categorical DOA and regression on spherical DOA.





\section{Proposed Method}
\begin{figure}[t]
\centering \includegraphics[scale=0.15]{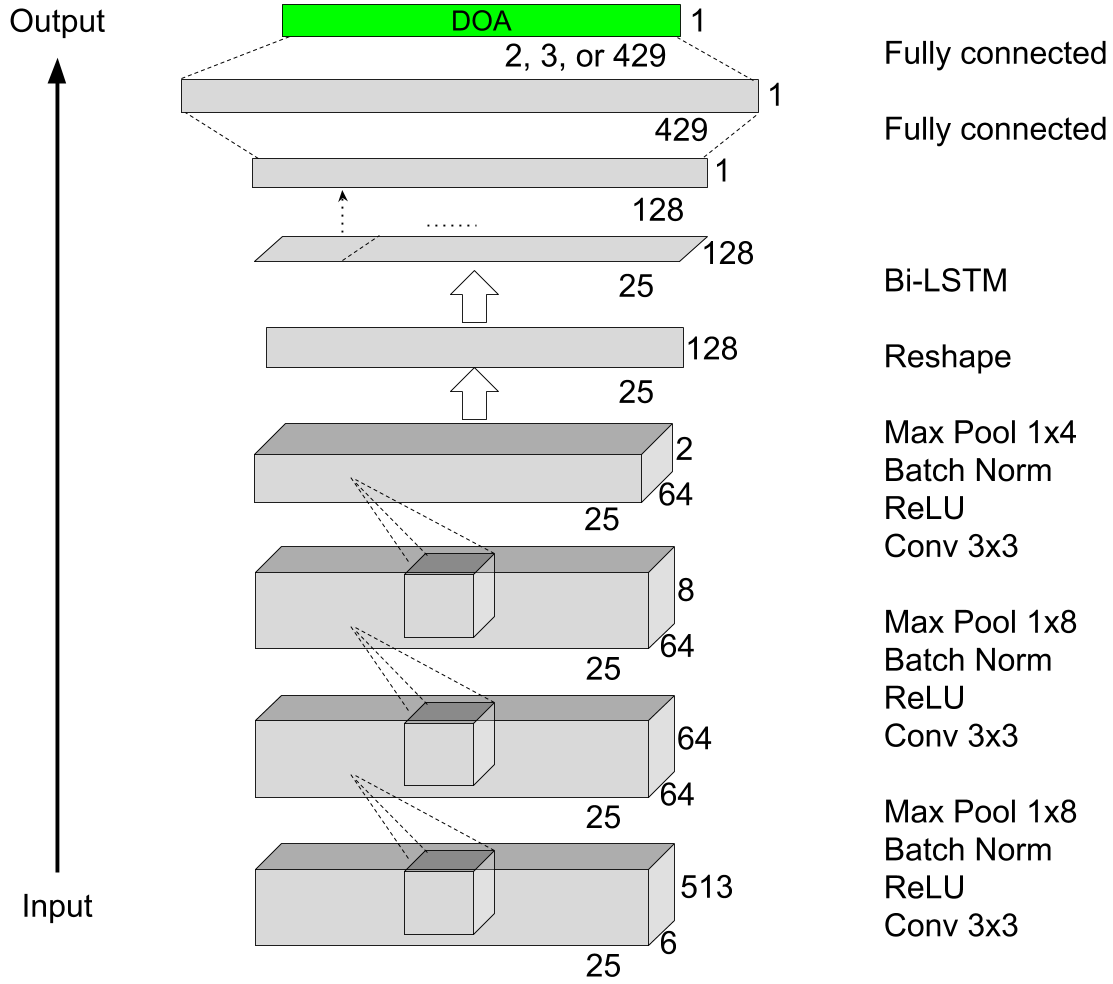}
\vspace{-1em}
 \caption{General network architecture for each regression network and classification network. The dimensionality of the output vector (shown in green) is 2 for spherical formulation, 3 for Cartesian, and 429 for classification. Note that our implementation of the classifier is equivalent to the implementation in~\cite{perotin2018crnn}, but our regression networks differ in the size of the output layer, and use 36\% fewer trainable parameters.}
 \vspace{-2em}
\label{fig:architecture}
\end{figure}

\subsection{Data Preparation}
Our DOA estimation network also relies on a large amount of labeled training data. However, collecting Ambisonic recordings and manually labeling them for training is tedious and time-consuming. Therefore, in speech/audio related training, the common practice is to use image-source methods to generate synthetic impulse responses for augmenting the training data~\cite{ko2017study}. However, the distribution of synthetic data may not match that of the real data well enough, which can cause large generalization error when applying a synthetically trained DNN to real test data. To overcome this issue of domain mismatch, a more accurate approach for generating training data is needed.

Sound propagation methods compute the reflection and diffraction paths from the sound sources to a listener in the virtual environment. Image-source methods do not model sound scattering or diffuse reflections, which are important phenomenons in acoustic environments. We utilize the state-of-the-art geometric sound propagation method~\cite{schissler2011gsound,schissler2014high,schissler2017interactive} to generate synthetic data that is more accurate than image-source methods. Its benefit has also been observed in other speech tasks~\cite{tang2019improving}.

Following the suggested procedure in~\cite{perotin2018crnn}, we generated 42,000 rectangular room configurations with dimensions uniformly and independently sampled between $2.5m\times 2.5m\times 2m$ and $10m\times 10m\times 3m$. Under each room configuration, we randomly populate three paired source-listener locations, both at least $0.5m$ away from walls. The geometric sound propagation method based on path tracing is used on each source-listener pair to generate its spatial room impulse response (SRIR). Then we convolve each SRIR with a randomly selected one-second clean speech sample from the Libri ASR corpus~\cite{panayotov2015librispeech} to generate realistic reverberant speech recordings in Ambisonic format. Babble and speech shaped noise~\cite{valentini2017noisy} are added to the convolved sound at signal-to-noise ratios (SNRs) following a normal distribution centered at $15dB$ with a standard deviation of $1dB$ as recommended by~\cite{yin2015noisy}. A short time fourier transform (STFT) is used to convert speech waveforms to spectrogram, and the features are extracted according to Section~\ref{sec:feature}. 

\subsection{Ambisonic Input Features}
\label{sec:feature}

In theory, Ambisonics of an infinite number of bases can reproduce the recorded soundfield with no error. As a practical approximation, we use the first four bases/channels necessary for first-order ambisonics (FOA). The FOA channels are denoted by $W, X, Y, Z$, where $W$ channel contains the zeroth-order coefficients that represents the omnidirectional signal intensity, and $X, Y, Z$ channels contain the first-order coefficients that encode direction modulated information. For a plane wave with azimuth $\theta$ and elevation $\phi$ creating a sound pressure $p$, the complex FOA components are:
\begin{gather}
\begin{bmatrix}
W(t,f) \\
X(t,f) \\
Y(t,f) \\
Z(t,f)
\end{bmatrix}
=
\begin{bmatrix}
1 \\
\sqrt{3}\text{cos}\theta \text{cos}\phi\\
\sqrt{3}\text{sin}\theta \text{cos}\phi \\
\sqrt{3}\text{sin}\phi
\end{bmatrix}
p(t,f),
\label{eq:foa}
\end{gather}
where $t$ and $f$ are time and frequency bins. We follow the approach in~\cite{perotin2018crnn} to construct input features from raw FOA audio. The active and reactive intensity vectors are encoded as:
\begin{gather}
\boldI(t,f)=
\begin{bmatrix}
W(t,f)*X(t,f) \\
W(t,f)*Y(t,f) \\
W(t,f)*Z(t,f)
\end{bmatrix},\\
\boldI_a(t,f)=\mathcal{R}\{\boldI(t,f)\},\boldI_r(t,f)=\mathcal{I}\{\boldI(t,f)\},
\label{eq:intensity_vec}
\end{gather}
where $\mathcal{R}\{\cdot\}$ and $\mathcal{I}\{\cdot\}$ extract the real and imaginary components of a complex signal respectively. Both feature vectors are divided by $|W(t,f)|^2+\frac{1}{3}(|X(t,f)|^2+|Y(t,f)|^2+|Z(t,f)|^2)$
to have a uniform range for deep neural network training.


\subsection{Output and Loss Formulation}
\label{sec:formulation}
We compare Cartesian $(x,y,z)$ and Spherical $(\theta, \phi)$ output representations in addition to the common Categorical representation in this work. A stacked CRNN formulation using Categorical outputs has been proposed in~\cite{perotin2018crnn}. We use this network structure and derive a set of Categorical, Cartesian, and Spherical forms from it, differing only in the size of the output layer, our independent variable. Maintaining a high similarity among the network architecture enables us to conduct well controlled comparisons between output representations for DOA estimation. We visualize our network architecture for the following three output formulations in Fig.~\ref{fig:architecture}. Key differences between the formulations are summarized in Tab.~\ref{tab:output}.

\subsubsection{Categorical Outputs}
We discretize the continuous DOA space into some number of possible categorical outputs. The angular resolution is chosen to be 10$\degree$, which results in 429 direction classes. Each training DOA label is assigned to the direction class that has the smallest angular difference from itself. We use this labeling to train a classifier that outputs a sigmoid vector of the direction class. The model is trained using cross-entropy loss.

\subsubsection{Cartesian Outputs}
We define the output of the Cartesian network as a 3-D vector, representing DOA in Cartesian coordinates. Training labels are unit vectors in $\mathbb{R}^3$ pointing toward the source. We use mean-squared error (MSE) as the loss to train our networks. Note that the output of the network is not constrained to lie on the unit sphere. As a result, a hypothetical output that is a large scalar multiple of the DOA label will result in a large loss, despite the perfect alignment of the output vector with the label. In practice, this property does not prevent our formulation from generating an accurate predictor, as shown in Section~\ref{sec:results}.

\subsubsection{Spherical Outputs}
In contrast to the Cartesian formulation, the spherical formulation encodes DOA as a 2-D vector representation azimuth ($\theta$) and elevation ($\phi$) angles. This representation has only 2 degrees of freedom in 3-D, which means the Cartesian representation has added a redundant dimension to this learning problem. One issue with using this form is that the periodicity of spherical angles makes distance computation between two angles more complicated than in Cartesian coordinates. Conventional mean squared loss is discontinuous, and therefore non-differentiable, over predicted azimuth and elevation, which eliminates the guarantee of convergence of Gradient Descent. Instead, we compute the great-circle distance on a sphere's surface using the haversine formula, which is differentiable:
\begin{align}
\label{eq:haversine}
h&=\sin^2\left(\frac{\phi_2 - \phi_1}{2}\right) + \cos(\phi_1) \cos(\phi_2)\sin^2\left(\frac{\theta_2 - \theta_1}{2}\right), \nonumber\\
D&=2r\text{arcsin}(\sqrt{h}),
\end{align}
where $r$ is the radius of the sphere, and $D$ is the great-circle distance between azimuth-elevation angles $(\theta_1, \phi_1)$ and $(\theta_2, \phi_2)$. By setting $r=1$, we define $D$ as our \emph{Haversine loss}.

\begin{table}[htbp]
\caption{Three types of output representations studied in this paper. $d$ is the number of classes into which a spherical surface is discretized. In our experiment, $d=429$.}
\vspace{-1em}
\label{tab:output}
\centering
\begin{tabular}{cccc}
\toprule
Output Type   & Activation &  Dimension & Loss Function \\\hline
Categorical  & Sigmoid  &  $\mathbb{R}^d$ & Cross-entropy \\
Cartesian &  Linear &  $\mathbb{R}^3$ & MSE \\
Spherical   & Linear & $\mathbb{R}^2$ & Haversine 
\\\bottomrule
\end{tabular}
\vspace{-2em}
\end{table}

\subsection{Model Architecture and Training Procedure}

As described in Section~\ref{sec:feature}, the inputs to these networks are active and reactive intensity vectors. The size of the input is 6x25x513, containing 513 frequency bins (16kHz sample rate), for each 6-D intensity vector, computed at each of 25 frames.

The CRNN architecture is shown in Fig.~\ref{fig:architecture}. There are three convolutional layers, each consisting of 2-D convolution, a rectified linear unit (ReLU), batch normalization, and max pooling. The outputs of each of the three layers are 64x25x64, 64x25x8, and 64x25x2, respectively. The output of the last layer is flattened to a 128-D vector for each of the 25 frames. Each frame's vector is fed into a two-layer bi-directional LSTM, and the output of the LSTM for each frame is fed through two time-distributed, fully-connected linear layers, generating a DOA estimate for each frame. As described in Section~\ref{sec:formulation}, we generate three forms of outputs depending on implementation: 2-D azimuth-elevation angles, 3-D Cartesian coordinates, or 429 DOA classes. During training, losses are computed at each frame and error is backpropagated through the network. During evaluation, a single DOA estimate is taken as the uniformly weighted average of all estimates across frames.


\section{Experiment and Results}
\label{sec:results}
\subsection{Benchmarks}

\begin{figure*}[htbp]
    \centering
    \subfigure[Recording 1]{
        \includegraphics[width=0.32\linewidth]{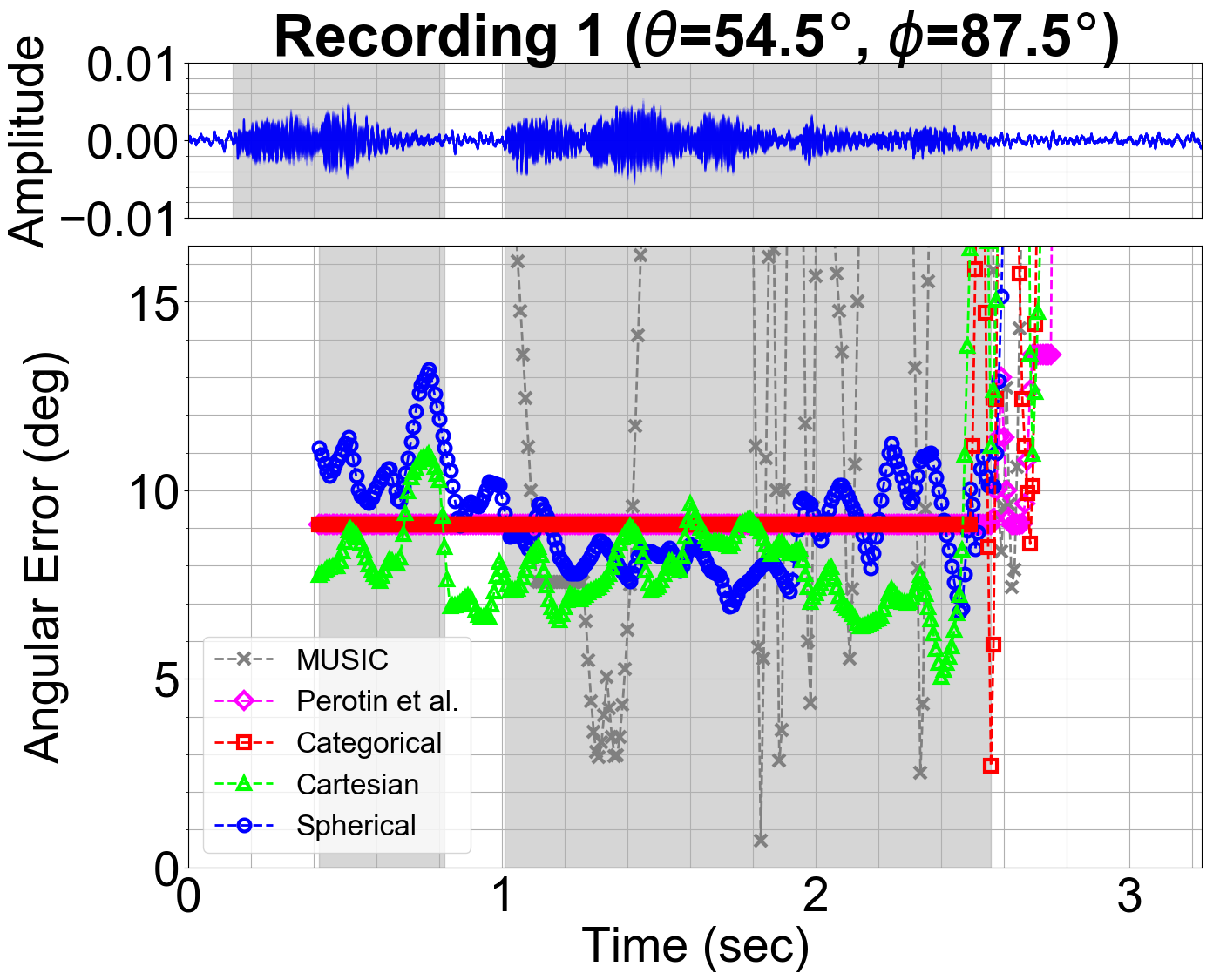}
        \label{fig:rec1}
    }
    \subfigure[Recording 2]{
        \includegraphics[width=0.32\linewidth]{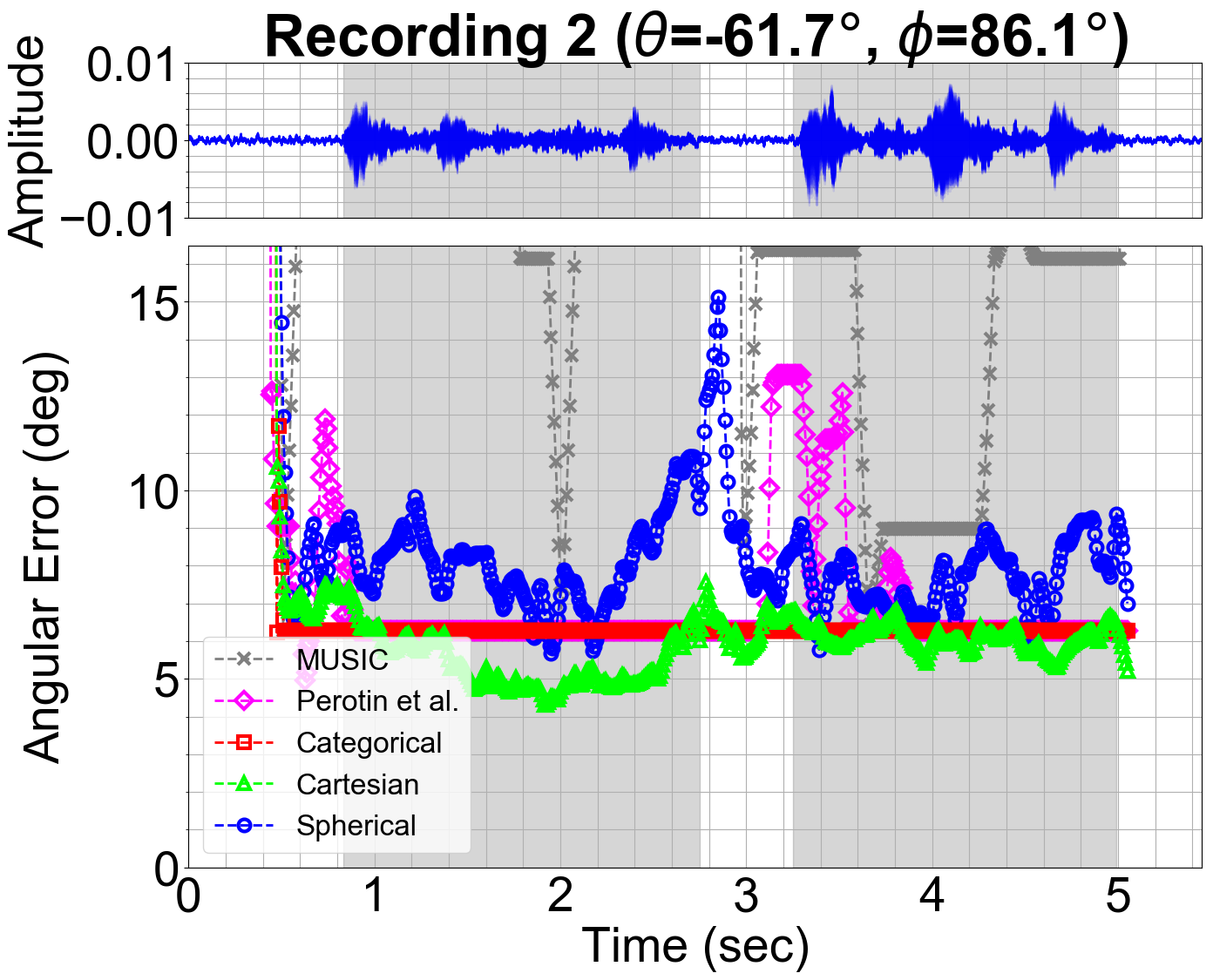}
        \label{fig:rec2}
    }
    \subfigure[Recording 3]{
        \includegraphics[width=0.32\linewidth]{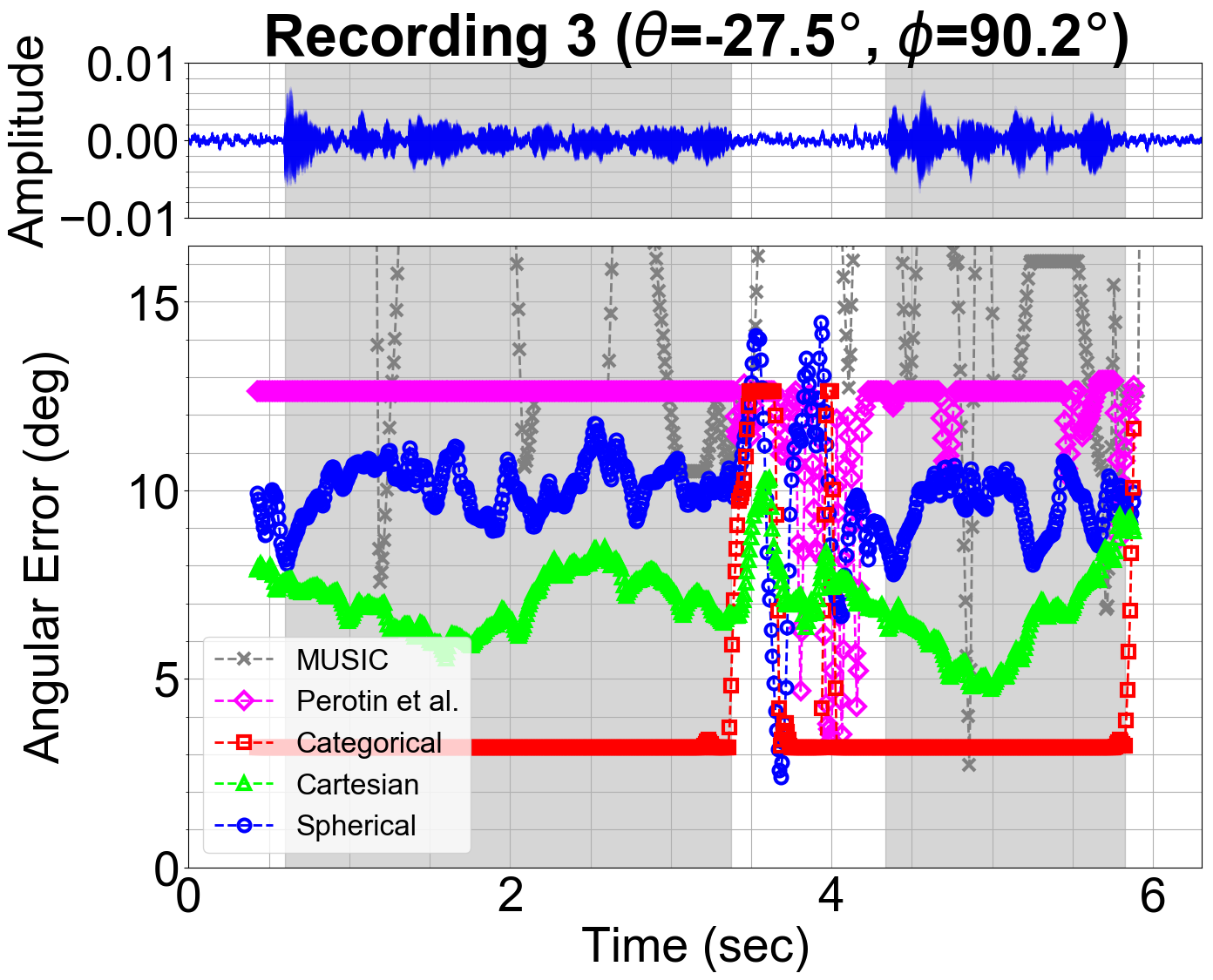}
        \label{fig:rec3}
    }
    \vspace{-1.5em}
\caption{Waveforms (top row) and angular tracking error (bottom row) for Recordings 1-3 in LOCATA Task 1. Shaded regions in the waveform indicate voice-active regions, while shaded regions in angular tracking error indicate the intersection of voice activity and regions containing predictions for all models. Each model must wait for a complete input window to make a prediction, hence the regions are not always identical between waveform and tracking error. Cartesian and Categorical models trained on our synthetic dataset achieve consistently lower tracking error compared with the classifier trained by Perotin et al.~\cite{perotin2018crnn} and the MUSIC algorithm~\cite{schmidt1986music}.
}
\vspace{-1.5em}
\label{fig:locata_res}
\end{figure*}

We evaluate each model on 1,189 samples from three static-source microphone signals in the third-party sound localization and tracking (LOCATA) challenge dataset~\cite{LOCATA2018a}, and a dataset of ambisonic RIRs accompanying the Spatially Oriented Format for Acoustics (SOFA) convention~\cite{perez2018ambisonics}. No real RIRs are involved during the training phase. The noise and clean speech used for training and test are from different datasets. We use Eq.~(\ref{eq:haversine}) as our angular error metric for visualization in Fig.~\ref{fig:locata_res}. 

Each signal in the LOCATA dataset is a real-world ambisonic speech recording with optically tracked azimuth-elevation labels. In theory, ambisonic coefficients up to the 4th order can be captured by an Eigenmike microphone, but we only use its first order components.

From the SOFA dataset, we extract 225 SRIRs recorded in the Alte Pinakothek museum using Eigenmike\textregistered, Sennheiser AMBEO\textregistered, and SoundField\textregistered microphones. Positions and rotations of all loudspeaker and microphones are provided by measurements using laser meter and pointers. A reverberant test set is generated by convolving each SOFA SRIR with a random 1-second clip from the CMU Arctic speech databases~\cite{kominek2004cmu}. Recorded background noise from the LOCATA dataset is added at a mean SNR level of $10dB$ with a standard deviation of $5dB$.

\subsection{Results and Analysis}
\subsubsection{LOCATA Dataset}
We compute average error along the temporal axis, for each static-source signal in LOCATA Task 1. Estimates of DOA are generated for each frame in the microphone signal using a sliding window. The resulting estimates are interpolated to the timestamps provided in the LOCATA dataset. Prediction error is computed as the angular distance between the prediction and the ground-truth DOA. Angular tracking error is visualized in Fig.~\ref{fig:locata_res}, for each predictor. Average angular error is computed over 234, 439, and 512 timestamps for Recordings 1, 2, and 3, respectively. Timestamps are selected to compute angular error if each algorithm makes a prediction for that timestamp, and the timestamp is located inside a voice-active region.

To generate a prediction at frame $i$ using a neural network, we feed in the sequence of 25 frames centered at $i$, and generate a sequence of 25 outputs of the network. If the output is Cartesian or spherical, we estimate DOA at $i$ as the average of the outputs at each frame. If the output is a classification grid, we average the output grid over the frames, to produce a cumulative score for each DOA, and choose the DOA with highest score as the prediction. When running the MUSIC algorithm, we restrict it to 4-channel recordings, as well, for fair comparisons.


\begin{table}[htbp]
\caption{Average angular tracking error within voice-active regions of LOCATA Task 1 recordings. Best performance in each column is highlighted in \textbf{bold}. All models are trained on data using specular and diffuse reflections in the geometric propagation algorithm, except for the MUSIC algorithm, which does not rely on training data. Perotin et al.~\cite{perotin2018crnn} refers to the Categorical model trained on data generated by the image-source method.}
\vspace{-1em}
\label{tab:locata}
\centering
\begin{tabular}{ccccccc}
\toprule
Model & Recording 1 & Recording 2 & Recording 3   \\\hline
MUSIC  &  18.6\textdegree & 16.9\textdegree & 17.5\textdegree  \\
Perotin et al.  & 9.1\textdegree  &  6.7\textdegree & 12.5\textdegree  \\
Categorical  & 9.3\textdegree & 6.3\textdegree & \textbf{3.2\textdegree} \\
Cartesian   & \textbf{8.5\textdegree} & \textbf{5.8\textdegree} & 6.8\textdegree \\
Spherical   & 9.2\textdegree & 7.9\textdegree & 9.9\textdegree
\\\bottomrule
\end{tabular}
\end{table}
\vspace{-1em}

We observe that in Fig.~\ref{fig:locata_res} and Tab.~\ref{tab:locata}, our Cartesian model achieves consistently lowest error on Recording 1 and 2, while our categorical model shows best performance in Recording 3. However, the Spherical model yields higher error than the other two models. We also observe that re-training the original model from Perotin et al.~\cite{perotin2018crnn} using data generated by geometric method to augment data results in lower tracking error compared with the use of the image-source method.

\subsubsection{SOFA Dataset}

\begin{table}[hbtp]
\caption{Results on the SOFA dataset. First three columns show the percentage of DOA labels correctly predicted within error tolerances, followed by average angular errors, and \%-improvement on baseline. Best performance in each column is highlighted in \textbf{bold}.}
\label{tab:sofa}
\centering
\begin{tabular}{cccccc}
\toprule
Model   & $<5$\textdegree &  $<10$\textdegree & $<15$\textdegree & Error & Improv. \\\hline
Perotin et al. & 11.9\%  & 35.9\%  &  73.2\%   & 16.9\textdegree & -\\
Categorical  & \textbf{24.4\%} & 58.2\% & \textbf{88.7\%} & 9.96\textdegree & 41\% \\
Cartesian &  \textbf{24.4\%} & \textbf{66.3\%} & \textbf{88.2\%} & \textbf{9.68\textdegree} & \textbf{43\%}\\
Spherical   & 18.2\% & 55.8\% & 82.5\%  &  11.2\textdegree  & 34\%
\\\bottomrule
\end{tabular}
\vspace{-1em}
\end{table}

A larger scale test is performed using the SOFA dataset. MUSIC algorithm is not evaluated because this dataset does not provide microphone hardware configurations required by MUSIC. We compute the percentage of correctly predicted directions under error tolerances of 5\textdegree, 10\textdegree, and 15\textdegree, as well as each model's average angular error on the whole dataset. It can be seen from Tab.~\ref{tab:sofa} that our Cartesian model consistently achieves the best performance under each column, outperforming the baseline model by 43\% in terms of average prediction error. 

During our training procedure, we notice that each model is able to converge within dozens of epochs. However, the number of trainable parameters in regression models (i.e. Cartesian and Spherical) is only 64\% of that in the classification model. This suggests that regression models tend to have a hypothesis set with lower complexity, which results in lower generalization error when tested on real data. We also tested the option of letting all models have approximately the same amount of trainable parameters, which has degraded the performance of the regression models. In conclusion, we are able to train a regression model that has superior performance over its corresponding classification model, although we do not observe obvious benefits in using the Spherical formulation.
\section{Discussion and Future Work}
In this paper, we demonstrate the benefits of using a geometric sound propagation simulator, as compared with image source methods for training DOA estimation networks, by reporting a higher accuracy on evaluation data. We evaluate the performance of a CRNN model in three output formulations: categorical, Cartesian, and spherical. We test them on two 3rd-party datasets and show that our Cartesian regression model achieves superior performance over classification and spherical models.

Evaluating classification models involves an additional factor: the resolution of the classification grid, which we kept fixed. Further, our work is limited to single-source localization problems, whereas in multi-source localization problems, classification models may have intrinsic advantages over regression models. Lastly, we restricted our simulation to extremely simple room settings to guarantee a fair comparison with the image-source method. Furture work may involve experimentation on more complex room configurations.



\bibliographystyle{IEEEtran}

\bibliography{mybib}

\end{document}